\documentclass[amsfonts,amssymb,amsmath,eqsecnum]{revtex4-1}

\usepackage{bm}
\usepackage{graphicx}
\usepackage{color}

\newcommand{\stl}[1]{ \mbox{ $ \hspace{0.1em} \stackrel{ \rule{0.3pt}{0.275ex} \hspace{-0.25pt} \overline{\hphantom{\mbox{$\displaystyle #1$}}} \hspace{-0.30pt} \rule{0.3pt}{0.275ex} }{#1} \hspace{0.2em} $ } } 

\newcommand{\abs}[1]{\left\vert#1\right\vert}

\begin{document}
\title{Oscillatory motion of sheared nanorods beyond the nematic phase}
\author{David~A.~Strehober}
\email{physik@strehober.de}
\author{Harald Engel}
\author{Sabine~H.~L. Klapp}
\email{klapp@physik.tu-berlin.de}
\affiliation{Institut f\"ur Theoretische Physik, Sekr. EW 7-1, Technische Universit\"at Berlin,
Hardenbergstra{\ss}e 36, D-10623 Berlin, Germany}

\begin{abstract}
We study the role of the control parameter triggering nematic order (temperature or concentration) on
the dynamical behavior of a system of nanorods under shear. Our study is based on a set of mesoscopic
equations of motion for the components of the tensorial orientational order parameter.
We investigating these equations via a systematic bifurcation analysis 
based on a numerical continuation technique, focusing on spatially homogeneous states. 
Exploring a wide range of parameters we find, unexpectedly, that states with oscillatory motion can exist even
under conditions where the equilibrium system is isotropic.
These oscillatory states are characterized by wagging motion of the paranematic director, and they occur
if the tumbling parameter is sufficiently small. We also present full non-equilibrium
phase diagrams, in the plane spanned by the concentration and the shear rate.
\end{abstract}

\maketitle
\section{Introduction}
The non-equilibrium dynamics of fluids composed of interacting, rod-like particles under external shear flow is a long-standing problem receiving
continuous attention both from a fundamental perspective and in applications \cite{GompperSchickVol2,deGennes,Olmsted1999,Fielding2003}. Renewed theoretical interest results from the discovery of stable, 
oscillatory orientational modes as well as rheochaos under conditions where the particles strongly interact \cite{Tao2005,Forest2007}. This is typically the case at high 
densities and/or low temperatures, depending on whether the system is rather lyotropic (density-driven) or thermotropic (temperature-driven). 
In thermal equilibrium, strong coupling between the particles leads to various mesophases with long-range orientational ordering and different degrees 
of positional order \cite{Fraden2000,Fraden2004}. Under shear, one observes not only shifts of these various transitions \cite{Olmsted92}, but
also oscillatory motion (and other dynamical modes) of the entire nematic director. Such behavior is predicted, e.g. by various continuum
theories (such as those of Doi and Hess) \cite{Hess1975,Hess1976,Hess1976b,Doi81} for the dynamics of the orientational order parameter
of the system, as well as in particle-based computer simulated techniques such as (event-driven) Brownian dynamics and multi particle collision
dynamics \cite{Tao2009,Ripoll2008a,Ripoll2008}. Experimentally, oscillatory states have been seen, e.g., in suspensions of fd-viruses in plane 
Couette flow geometries \cite{Lettinga2004,Lettinga05}. Again, the corresponding equilibrium system (at zero shear rate) is a nematic.

However, despite the large number of studies already carried out, there are still some rather fundamental questions left. One of these is whether
oscillatory motion can occur {\em only if} the system in the stable or metastable nematic regime, as it is often assumed. In other words, is it 
possible that spontaneous oscillatory motion develops on the isotropic side of the shear-distorted phase diagram? And if such oscillations indeed exist,
how do they transform into other dynamical states if the shear rate or other system parameters are changed? Or, in the language of non-linear dynamics,
what is the nature of the underlying bifurcations?

In the present paper we aim to shed light on these issues on the basis of the mesocopic approach suggested by Hess \cite{Hess1975,Hess1976,Hess1976b}
and later used in many subsequent studies \cite{GrandnerEPJ2007,GrandnerPRE2007,Heidenreich2006,Heidenreich2009}. 
However, contrary to these earlier studies we here employ a systematic bifurcation analysis, which allows us to explore a much wider range of 
parameters and, in addition, to fully characterize the nature of the dynamical transitions.

In the Hess-Doi approach, the 
relevant dynamic variable is the tensorial orientational order parameter whose dynamics is determined by the (Couette) shear, on the one hand, and relaxational contributions
(derived from a Landau free energy), on the other hand. By using a suitable tensor basis, the corresponding equation of motion
transforms into a set of five, non-linear first-order differential equations for the expansion coefficients.  It is well established that these equations predict rich state diagrams
involving a variety of oscillatory states and even chaos. Here we analyze the dynamical behavior on basis of a numerical path continuation method, exploring two bifurcation parameters, that is,
the shear rate and a coupling constant). Previous works by Alonso \emph{et~al.} \cite{Alonso2003} and Forest \emph{et~al.} \cite{Forest2004} also used
 bifurcation analysis. However, they limited the description to orientations within the shear plane \cite{Alonso2003} or focused on the behavior for weak shear \cite{Forest2004}.
Another bifurcation analysis was presented by Rey \cite{Rey1995}, who considered (contrary to our work) a system under steady biaxial stretching flow.\\
Here we consider a steady \emph{uniaxial} shear flow, but do not restrict the resulting orientational dynamics.
Our bifurcation analysis leads to results that indicate there is indeed an "island" in parameter space where the equilibrium system is a stable isotropic state, whereas shear induces an oscillatory (wagging) motion. Moreover, we can extract entire non-equilibrium state diagrams revealing a close relationship to experimental results.

The paper is organized as follows. We first we give an overview of the model system and the governing dynamical equations. Next we apply the bifurcation analysis
to a specific state parameter within the nematic phase, the main purpose being to compare with earlier results obtained by simple numerical integration.  We then move towards
previously unexplored parameter regions and set up full non-equilibrium diagrams. Finally, we give a conclusion and an outlook. 
The paper is rounded up by two appendices describing details of our method(s).

\section{Theory}
\subsection{Second rank alignment tensor}
We are interested in the dynamics of a suspension of rodlike particles under shear. 
As a method of investigation we employ a mesoscopic approach originally suggested by Doi and Hess \cite{Hess1975,Hess1976,Hess1976b,Doi80,Doi81} where
the basic dynamical variable is the orientational order parameter $ \bm{\tilde{a}}(\bm{r},t)$ (in the literature also often $\bm{Q}$). The latter describes
the average orientation of the rods in a small, yet macroscopic volume of space. The alignment tensor is linked to the electric susceptibility tensor \cite{Sonnet1995}, and can be measured optically
by birefringence. In an unsheared (i.e., equilibrium) nematic system it is usually sufficient to describe the orientational order by a vector, the so-called director, which indicates the average direction of the particles.
The strength of this average orientation is then described by the Maier-Saupe parameter $S$. However, in the presence of shear the orientational order can become
biaxial \cite{Rien02,GrandnerEPJ2007}. Therefore, the order parameter $\bm{\tilde{a}}(\bm{r},t)$ must be a second-rank tensor, which we define according to
\begin{equation}
 \label{eq:alignment}
 \bm{\tilde{a}}(\bm{r},t) = 
 \sqrt{\frac{15}{2}} \langle \bm{u}\otimes\bm{u} -\frac{1}{3}\mathrm{Tr}(\bm{u}\otimes\bm{u})\rangle_{\mathrm{or}} 
  \equiv\sqrt{\frac{15}{2}} \langle\stl{\bm{u}\otimes\bm{u}}\rangle_{\mathrm{or}},
\end{equation}
where the unit vector $\bm{u}=\bm{u}(\bm{r},t)$ stands for the orientation of the rod at position $\bm{r}$ at time $t$, 
$\otimes$ denotes the dyadic product, and $\mathrm{Tr}(\ldots)$ denotes the trace of a matrix. As it follows from the first member of Eq.~(\ref{eq:alignment}),
the tensor $\bm{\tilde{a}}(\bm{r},t)$ is symmetric and traceless (which we henceforth indicate by using the symbol $\stl{...}$).
Moreover, the average $\langle\ldots\rangle_{\mathrm{or}}$ appearing in Eq.~(\ref{eq:alignment}) is defined as
\begin{equation}
\label{eq:average}
\langle\ldots\rangle_{\mathrm{or}}= \int_{S^2}{\mathrm{d}}^2\bm{u}\,\ldots \,\rho^{\mathrm{or}}(\bm{u},\bm{r},t),
\end{equation}
involving the orientational distribution function $\rho^{\mathrm{o}r}(\bm{u},\bm{r},t)$ \cite{Sonnet1995}. The integral in Eq.~(\ref{eq:average})
is performed with respect to the angles describing the vector $\bm{u}$. The orientational distribution is defined as
$\rho^{\mathrm{or}}(\bm{u},\bm{r},t) = N^{-1} \langle \sum_{i=1}^N \delta(\bm{u}-\bm{u}_i(t))\rangle_{\mathrm{ens}}$, where $\bm{u}_i$ is the microscopic orientation of particle $i$ ($i=1,\ldots,N$),
and $\langle\ldots\rangle_{\mathrm{ens}}$ is an ensemble average 
in a small volume $dV$ around the space point $\bm{r}$ at time $t$.
Combining Eq.~(\ref{eq:average}) with the last member of Eq.~(\ref{eq:alignment}), we see that the order parameter tensor $\bm{\tilde{a}}(\bm{r},t)$ corresponds to the
second moment of $\rho^{\mathrm{o}r}(\bm{u},\bm{r},t)$. In other words, the orientational distribution function provides the link between the mesoscopic quantity
$\bm{\tilde{a}}(\bm{r},t)$ and the microscopic degrees of freedom, $\bm{u}_i$. 

The eigenvalues of the tensor $\bm{\tilde{a}}$ are $\mu_1$,$\mu_2$,$\mu_3$ and they are related to the eigenvectors ("principal axes")
$\bm{l}$,$\bm{m}$,$\bm{n}$. The tensor $\bm{\tilde{a}}$ can then be written in the form $\bm{\tilde{a}}= \mu_1 \bm{l}\otimes\bm{l}+\mu_2 \bm{m}\otimes\bm{m}+\mu_3 \bm{n}\otimes\bm{n} $ \cite{Rienacker2000}. The principal axis related to the largest eigenvalue is considered as the {\it director} of the system. Commonly, one assumes that this role is played
by $\mu_3$, such that the director is given by $\bm{n}$. For {\it uniaxial} order,
$\bm{n}$ is the only relevant axis, and $\mu_3$ is linked to the Maier-Saupe parameter $S$ via $\mu_3=2\sqrt{5/6}S$ \cite{Heidenreich2008Thesis}, 
where
$S\equiv \langle P_2({\bm{u}\cdot\bm{n}}) \rangle_{\mathrm{or}}$ and $P_2$ denotes the second Legendre polynomial.
For the more general case of {\it biaxial} order,
the principal axes can be conveniently visualized using super-quadratic tensor glyphs \cite{Kindlmann2005,Kelly2006}.
%

\subsection{Equations of motion for the homogeneous sheared system}
\label{subsec:eqmotionhomogeneous}
In the absence of shear, the ordering behavior of the system is essentially controlled by either the concentration, $c$, or the temperature, $T$. 
Specifically, $c$ is the control variable
for lyotropic liquid crystals, and for most colloidal rods (e.g. fd-viruses \cite{Lettinga05}), whereas thermotropic liquid crystals are controlled by the temperature.
Nematic phases typically occur at high concentrations or low temperatures, respectively. In the present paper we
describe lyotropic and thermotropic systems on the same footing in terms of an effective, dimensionless "temperature" $\theta$ defined as
\begin{eqnarray}
\label{theta}
 \theta & = & \frac{1-T^*/T}{1-T^*/T_{\mathrm{K}}}\quad\mathrm{(thermotropic)},\nonumber\\
 \theta & = & \frac{1-c/c^{*}}{1-c_{\mathrm{K}}/c^*}\quad\mathrm{(lyotropic)}.
\end{eqnarray}
In Eqs.~(\ref{theta}), $T_{\mathrm{K}}$ and $c_{\mathrm{K}}$ correspond to the temperature and concentration where the isotropic and nematic phases coexist \cite{Sonnet1995,Rienacker2000}.
In other words, $T_{\mathrm{K}}$ and $c_{\mathrm{K}}$ define the location of the first-order isotropic-nematic transition of the system. These parameters are sometimes
also referred to as ''clearing point'' due to the related change of the optical properties of the system. From Eqs.~(\ref{theta}) it follows that 
at $T=T_{\mathrm{K}}$ ($c=c_{\mathrm{K}}$), the effective temperature $\theta=1$. On the other hand
$\theta=0$ at the "pseudocritical" temperature $T^{*}$ or pseudocritical concentration $c^{*}$. For $T<T^{*}$ ($c>c^{*}$) the isotropic phase is globally unstable.
Between the lower spinodal temperature and the clearing point temperature $T_K$ (clearing point concentration $c_K$), i.e. $0<\theta<1$, without flow the nematic phase is stable and the isotropic phase is metastable.
In the range between temperature $T_K$ (concentration $c_{\mathrm{K}}$) and $T_u$ ($c_u$), i.e. $1<\theta<9/8$,
the nematic phase in the absence of shear is metastable and the isotropic phase is stable. That is, $T_u$ ($c_u$) marks
the upper limit (lower limit) of the metastable nematic phase \cite{Rey2012}.
The stability of the two phases and thus, the equilibrium behavior of the system,
is determined by the (dimensionless) Landau-de Gennes (LG) free energy \cite{deGennes}
\begin{align}
\label{LG}
  {\Phi}(\bm{a}) ={\frac{{\theta}}{2}}
  \mathrm{Tr}(\bm{a}{\cdot}\bm{a})-{\sqrt{6}}
  \mathrm{Tr}(\bm{a}{\cdot}\bm{a}{\cdot}\bm{a})+{\frac{1}{2}}
  (\mathrm{Tr}(\bm{a}{\cdot}\bm{a}))^2,
\end{align}
where we have focused on spatially homogeneous states (i.e., no gradient terms). Also,
we have defined the rescaled order parameter
\begin{align}
 \label{eq_reduceda}
 \bm{a} \equiv \frac{\bm{\tilde{a}}}{a_{\mathrm{K}}},
\end{align}
where $a_{\mathrm{K}}$ is the value of the parameter $a$ defined as
\begin{equation}
\label{order_parameter}
a=\sqrt{\mathrm{Tr}(\bm{a}\cdot\bm{a})}.
\end{equation}
evaluated at coexistence, that is, at $T_{\mathrm{K}}$ or $c_{\mathrm{K}}$. We note that, for the special case of {\it uniaxial} orientational order, the parameter $a$ is proportional to the
Maier-Saupe order parameter, that is, $a=\sqrt{5}S$.

In non-equilibrium the LG free energy, or rather its derivative with respect to the order parameter,
governs the {\it relaxational} dynamics of the system. This derivative follows from Eq.~(\ref{LG}) as
\begin{align}
  {\Phi}'(\bm{a}) &\equiv \frac{\partial \Phi}{\partial \bm{a}}
  = {\theta}\bm{a}-3{\sqrt{6}}{\stl{\bm{a}\cdot\bm{a}}}+2
  \mathrm{Tr}(\bm{a}{\cdot}\bm{a}){\cdot}\bm{a}. \label{eq:LdGderivative}
\end{align}
In the presence of shear flow, the relaxational dynamics competes with the dynamics induced by flow field $\bm{v}(\bm{r})$.
The resulting terms entering the equation of motion for ${\bm{a}}(t)$ 
can be derived from a generalized Fokker-Planck equation \cite{Hess1976,Doi80,Doi81}
or, alternatively, from irreversible thermodynamics \cite{Hess1975}. In these equations, 
the influence of shear is captured by the symmetric and antisymmetric part of the velocity gradient
\begin{align}
  \bm{{\Gamma}} &\equiv
  {\frac{1}{2}}(({\nabla}\bm{v})^T+{\nabla}\bm{v}) \label{eq:Gamma}\\
  \bm{{\Omega}} &\equiv
  {\frac{1}{2}}(({\nabla}\bm{v})^T-{\nabla}\bm{v}),\label{eq:Omega}
\end{align}
where $\bm{\Gamma}$ and $\bm{\Omega}$ are the so-called deformation and vorticity, respectively.
Here we focus on the case of plane Couette flow characterized by a linear velocity profile
$\bm{v}={\dot{{\gamma}}} y \bm{e}^x$,
where $\dot{\gamma}$ is the shear rate and $\bm{e}^x$ is a unit vector. Thus, the shear plane is spanned by the x- and y-directions and the unit vector $\bm{e}^z$ is orthogonal to the shear plane.

Combining relaxational and shear-induced terms, the equation of motion for a homogeneous system (i.e., $\bm{a}(\bm{r},t)=\bm{a}(t)$) in reduced units 
reads \cite{GrandnerEPJ2007}
\begin{align}
  {\frac{\mathrm{d} \bm{a}}{\mathrm{d} t}}=2
  \stl{\bm{{\Omega}}{\cdot}\bm{a}}+2{\sigma}\stl{\bm{{\Gamma}}{\cdot}\bm{a}}-{\Phi}'(\bm{a})+{\sqrt{{\frac{3}{2}}}}{{\lambda}_{\mathrm{K}}}\bm{{\Gamma}}. \label{eq:tensorhomogeneous}
\end{align}
In Eq.~(\ref{eq:tensorhomogeneous}), the shear flow enters through the three terms involving the quantities $\bm{\Gamma}$ and $\bm{\Omega}$ defined in Eqs.(\ref{eq:Gamma})-(\ref{eq:Omega}), respectively. Specifically, the first
term of the right side of Eq.~(\ref{eq:tensorhomogeneous}) describes the impact of the flow vorticity $\bm{\Omega}$, while the second term couples the deformation rate $\bm{\Gamma}$ linearly to the alignment tensor. The third term represents the relaxational part determined by the LdG free energy [see Eq.~(\ref{eq:LdGderivative})].
Finally, the last term on the right side of Eq.~(\ref{eq:tensorhomogeneous}) involves the so-called tumbling parameter $\lambda_{\mathrm{K}}$, which determines the impact 
of the external perturbation due to the flow. This coupling parameter can be related to the axis ratio $q=L/D$ (with $L$ length, $D$ width) of the particles
via \cite{Hess1976} 
\begin{equation}
\label{tumb_shape}
 \lambda_{\mathrm{K}}=\frac{2}{\sqrt{5}a_{\mathrm{K}}}\,
 \frac{q^2-1}{q^2+1}.
 \end{equation}
Specifically, one has $\lambda_{\mathrm{K}}=0$ for spherical particles whereas $\lambda_{\mathrm{K}}>0$ for elongated particles. As an example
we briefly consider fd-virus particles which have been experimentally studied under shear in \cite{Lettinga05,Lettinga2004}.
These particles have a rather large aspect ratio of $q=L/D\approx130$ \cite{Purdy2003,Dogic2006}, such that the ratio $(q^2-1)/(q^2+1)\approx 1$. To determine the value of $a_{\mathrm{K}}$, we use the fact 
that typical values of the Maier-Saupe parameter at coexistence
are about $S\approx 0.6-0.7$ \cite{Purdy2003} and $a=\sqrt{5}S$. Inserting these values into Eq.~(\ref{tumb_shape}) we obtain 
$\lambda_{\mathrm{K}}^{\mathrm{fd}}\approx 0.6-0.7$.

It is common to transform the tensorial equation~(\ref{eq:tensorhomogeneous}) into a system of scalar equations \cite{KlappHess2010}. This is done by expanding $\bm{a}$
and the other tensors appearing in Eq.~(\ref{eq:tensorhomogeneous}) into a tensorial basis set (see, e.g., \cite{Rien02a}), yielding
\begin{align}
  \dot{a}_0= & -\phi_0 -{\frac{1}{3}}{\sqrt{3}}{\sigma}\dot{{\gamma}}a_2\notag\\
  \dot{a}_1= & -\phi_1 +\dot{{\gamma}}a_2\notag\\
  \dot{a}_2= & -\phi_2 -\dot{{\gamma}}a_1+{\frac{1}{2}}{\sqrt{3}}{\lambda}_K\dot{{\gamma}}-{\frac{1}{3}}{\sqrt{3}}{\sigma}\dot{{\gamma}}a_0\notag\\
  \dot{a}_3= & -\phi_3+{\frac{1}{2}}\dot{{\gamma}}({\sigma}+1)a_4\notag\\
  \dot{a}_4= & -\phi_4+{\frac{1}{2}}\dot{{\gamma}}({\sigma}-1)a_3\label{eq:HomSyst1},
\end{align}
where
\begin{align}
  \phi_0= &({\theta}-3a_0+2a^2)a_0+3(a^2_1+a^2_2)-{\frac{3}{2}}(a^2_3+a^2_4)\notag\\
  \phi_1= &({\theta}+6a_0+2a^2)a_1-{\frac{3}{2}}{\sqrt{3}}(a^2_3-a^2_4)\notag\\
  \phi_2= &({\theta}+6a_0+2a^2)a_2-3{\sqrt{3}}a_3a_4\notag\\
  \phi_3= &({\theta}-3a_0+2a^2)a_3-3{\sqrt{3}}(a_1a_3+a_2a_4)\notag\\
  \phi_4= &({\theta}-3a_0+2a^2)a_4-3{\sqrt{3}}(a_2a_3-a_1a_4)\label{eq:HomSyst2}.
\end{align}
In Eqs.~(\ref{eq:HomSyst2}), $a^2 =\sum_{i=0}^4 a^2_i$. In Appendix \ref{appA}, we present some general remarks concerning the structure of Eqs.~(\ref{eq:HomSyst1}) and (\ref{eq:HomSyst2})
and the consequences for the dynamic behavior of the system.

In previous studies \cite{Hess2004}
it has been shown, for the case of planar Couette flow, the parameter $\sigma$ has only minor importance. Therefore, we henceforth set $\sigma=0$. 

The coupled set of ordinary differential equations (ODEs) given in Eqs.~(\ref{eq:HomSyst1}) can be studied using methods from nonlinear dynamics.
Here we perform a bifurcation analysis using numerical continuation techniques, specifically, the software package MATCONT \cite{MATCONT}. 
Some details of this method can be found in Appendix \ref{appB}.
\section{Numerical Results}
\label{sec:NumericalResults}
The subsequent section is divided into three parts. In the first part (Sec.~\ref{subsec:NematicPhase}), we aim to validate the results of our bifurcation analysis by comparing with well-established results from direct numerical integration \cite{GrandnerEPJ2007}. To this end we consider the temperature $\theta=0$ and calculate a state diagram in
the plane spanned by $\lambda_{\mathrm{K}}$ and $\dot\gamma$. The temperature $\theta=0$ corresponds to the lower spinodal
temperature $T^{*}$ (upper spinodal concentration $c^{*}$). In other words, for $\theta<0$ without shear the nematic phase is the only stable phase, while the isotropic phase is (locally and globally) unstable.
\\
For $\theta=0$ a large body of results for this specific value
already exists \cite{GrandnerEPJ2007,GrandnerPRE2007,Heidenreich2008Thesis,Rienacker2000}. 
Beyond the pure numerical comparison, however, we also analyze in detail the {\it nature} of the observed bifurcations which are, so far, largely unknown.
In the second part (Sec.~\ref{subsec:HighT}) we apply the bifurcation analysis to systems 
at larger values of $\theta$ corresponding to the isotropic side of the equilibrium I-N transition [see text below Eqs.~(\ref{theta})]. 
In particular, we demonstrate the existence of oscillatory solutions. As a summary and overview of our results, we finally present in Sec.~\ref{overview} full non-equilibrium phase diagrams
(in the plane $\theta$-$\dot\gamma$ or $c$-$\dot\gamma$, respectively)
for selected values of the tumbling parameter. Later in this section we also provide codimension-1 diagrams that help to interpret the codimension-2 state diagrams.

\subsection{Nematic Phase}
\label{subsec:NematicPhase}

In this subsection we set $\theta = 0$, where the unsheared equilibrium system is in the nematic phase. 
Switching on the shear flow the system can be brought into a variety of time-dependent and steady states
characterized by a particular behavior of the order parameter $\bm{a}(t)$ \cite{Rien02,GrandnerEPJ2007}. 
Here we investigate the dynamics in the plane spanned by the tumbling parameter $\lambda_{\mathrm{K}}$ and the shear rate $\dot{\gamma}$.

\begin{figure}
\includegraphics[width=\columnwidth]{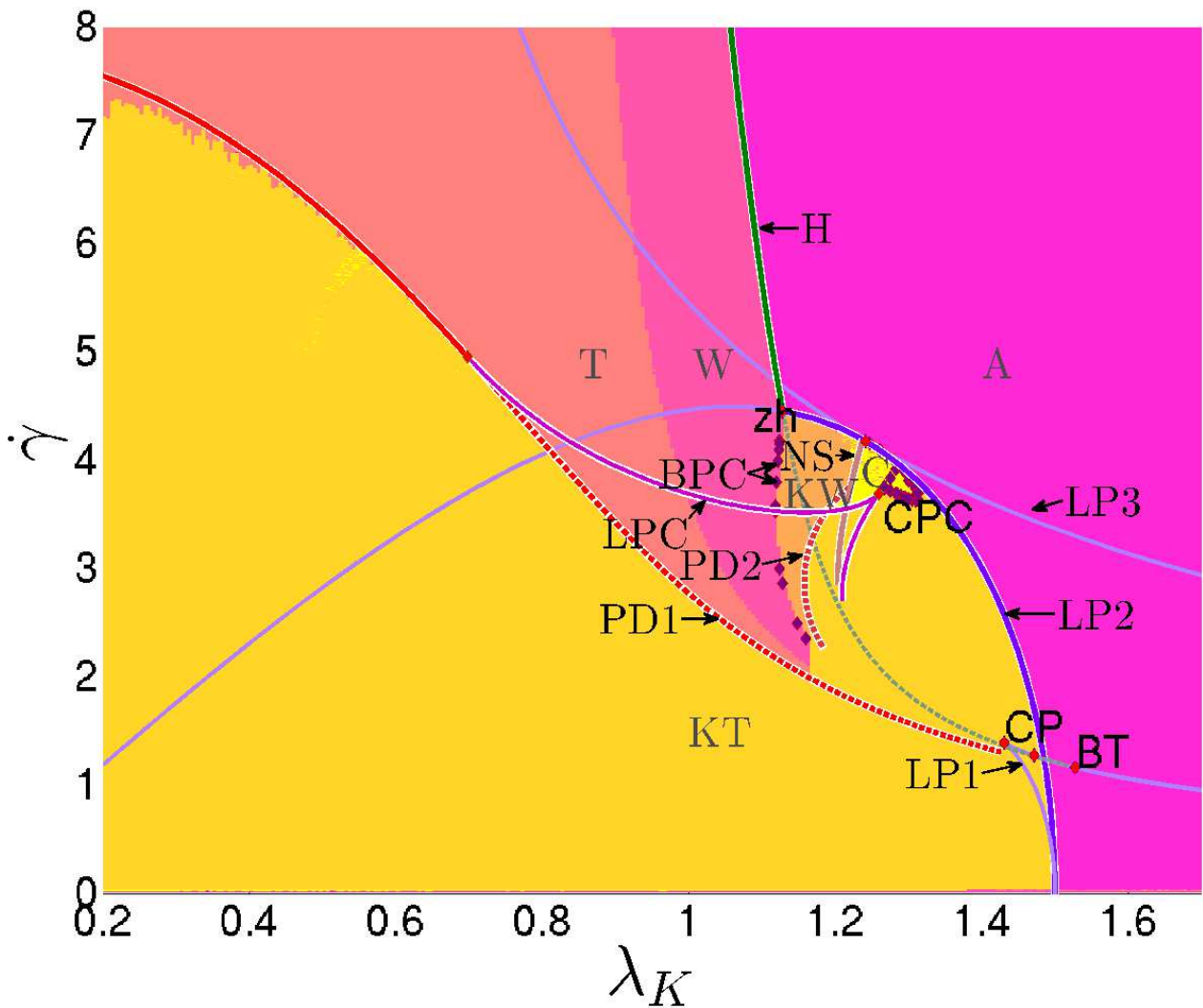}
\caption{\label{fig_maincomp_theta0}
(Color online) Dynamical state diagram in the plane spanned by the tumbling parameter $\lambda_{\mathrm{K}}$ and the shear rate $\dot\gamma$
at the reduced temperature $\theta=0$ (and $\sigma=0$). Areas are computed via direct numerical integration of the ODEs [see Eqs.~(\ref{eq:HomSyst1})
and (\ref{eq:HomSyst1})], whereas the curves are found from a codimension-2-bifurcation analysis. The areas are colored and labeled according to the dynamical state (A=Alignment, W=Wagging, T=Tumbling, KW=Kayaking Wagging, KT=Kayaking Tumbling, and C=Complex/Chaotic). The bifurcation branches are labeled as LP=Limit Point (saddle-node), H=Hopf, PD=Period Doubling, LPC=Limit Point of Cycles, BPC= Branch Point of Cycles, and NS = Neimarck Sacker. In addition, the diagram involves some special codimension-2 bifurcation points. These are the CP= Cusp Point, CPC= Cusp Point of Cycles, ZH= Zero Hopf, and BT= Bogdanov Takens point.}
\end{figure}

In Fig.~\ref{fig_maincomp_theta0} we show a composite of data obtained by direct numerical integration (colored areas), on the one hand, and the bifurcation analysis (curves), on the other hand. 
Within the calculations based on direct integration, we have considered a narrow grid of parameter pairs ($\lambda_{\mathrm{K}},\dot{\gamma}$). The resulting dynamics 
of the alignment tensor was then automatically classified (ignoring transient structures) and colored accordingly. The curves in Fig.~\ref{fig_maincomp_theta0}
were obtained using the codimension-2 bifurcation analysis described in the Appendix.
 
The different regions appearing in the phase diagram at $\theta=0$ are the
wagging state (W), the tumbling state (T), kayaking-wagging (KW), kayaking-tumbling (KT), and the (flow-)alignment state (A). Apart from the last
(stationary) state, where the director is "frozen" along a direction within the shear plane, all other states
mentioned so far are characterized by a time-dependent, oscillatory behavior 
of the coefficients $a_i(t)$ of $\bm{a}(t)$. Physically, these oscillations correspond to oscillations
of the nematic director either {\it within} the shear plane (W, T), or {\it out of} the
shear plane (KW, KT). In addition to these regular states, the diagram at $\theta=0$ also involves 
complex or chaotic (C) states, as has been shown in previous investigations \cite{GrandnerEPJ2007,Rien02}. 
In the present study, however, we will rather focus on the regular states.

We note that the colored areas in Fig.~\ref{fig_maincomp_theta0}, which have been obtained by direct integration, agree quantitatively with earlier numerical results
(see, e.g., \cite{GrandnerEPJ2007}). More importantly, however, we can see from Fig.~\ref{fig_maincomp_theta0} that these colored areas are closely bounded by the curves stemming from our bifurcation analysis.
This already indicates that the two approaches yield consistent results for the general dynamical behavior of the system. 
In the following we discuss in more detail the boundaries between the states, that is, the nature of the underlying bifurcations.

\subsubsection{From the steady state to in-plane oscillations}
We begin by considering the situation encountered at high shear rates and large tumbling parameters (see upper right corner of the diagram Fig.~\ref{fig_maincomp_theta0}). In this case, the system
settles into a "flow-aligned" (A) state where the nematic director remains temporarily fixed along a direction within the shear plane. 
From a physical point of view, this A state may be further characterized by the magnitude of the nematic ordering under shear and the flow angle $\varphi$ between director and flow direction. 
Mathematically, the A state
corresponds to a (stable) fixed point of the dynamical system
Eq.~(\ref{eq:HomSyst1}). In the case of $\theta=0$ considered in this Section, there is only one stable fixed point and thus, only one type of alignment. More generally, however, the A state can also
involve more than one fixed point. An example will be given in Sec.~\ref{subsec:HighT} where we consider a higher temperature $\theta$ and find, within the A state, a coexistence 
of paranematic and nematic flow alignment \footnote{We note in this context that our algorithm does not distinguish between presence of two coexisting shear-aligned states
and presence of only one such state.}.

Starting from the A state, we now decrease the tumbling parameter $\lambda_{\mathrm{K}}$. Provided that the shear rate is sufficiently large ($\dot{\gamma}\gtrsim 4.5$), the system
then encounters a dynamical transition into the wagging (W) state (see Fig.~\ref{fig_maincomp_theta0}). In the W state the angle $\varphi$
between the director (which still lies in the shear plane) and the flow direction oscillates periodically
between a minimal and a maximal value. This
angular motion is accompanied by an oscillation of the
magnitude of nematic order.

The dark green line (upper central dark line), labeled by H, which separates the A and the W region in Fig.~\ref{fig_maincomp_theta0}, denotes a Hopf bifurcation line. Indeed, within the {\it entire} regime left and below of the H line (i.e., not only in the W state), the system exhibits oscillatory dynamics corresponding to stable limit cycles (see also \cite{Alonso2003,Andrews1995,Faraoni1999,Maffettone2000}). Moreover, we find that the Hopf bifurcation separating A and W states is {\it supercritical}.
This implies that, upon crossing the H line, the W state is "born" with zero amplitude, but finite frequency at the onset. 

Lowering $\lambda_{\mathrm{K}}$ even further, the character of the oscillatory state (i.e., the limit cycle) changes from wagging to tumbling (T). This latter state involves again
a periodic motion where the director performs {\it full} rotations
rather than just wagging (W) in the shear plane. We stress, however, that there is no fundamental difference between W and T motion in the sense that
the change from W to T is not a true bifurcation (see also \cite{Alonso2003}). Finally, the curve corresponding to the lower boundary of the W/T domain (red line)
is a period doubling bifurcation curve (PD1). Below the PD1 curve, the W or T state do not correspond to stable limit cycles any more.
As seen in the very left of Fig.~\ref{fig_maincomp_theta0} (i.e., small $\lambda_{\mathrm{K}}$), there are small deviations
between the location of the PD1 curve (obtained by the bifurcation analysis) and the colored region of T states obtained by direct numerical integration.
These deviations arise from the fact that there are very long-living transients in this particular parameter region, making the classification of the final state based on the integration scheme
rather difficult. However, this "problem" could be fixed by extending the numerical integration towards even longer time scales.
\subsubsection{Emergence of symmetry-breaking states}
Starting from a W or T state (see upper left area of the diagram Fig.~\ref{fig_maincomp_theta0}) and reducing the shear rate $\dot{\gamma}$ we 
cross the PD1 curve, thereby entering  the domain of kayaking-tumbling (KT) states.
Within the KT state, the director performs (full) rotations {\it out of} the shear plane (contrary to the T state mentioned above). Thus, KT is an example of a
state breaking the symmetry introduced by the shear flow geometry.

The nature of the transition separating the W/T and the KT regime depends on the value of the tumbling parameter. This is indicated by the solid and dashed parts of the
PD1 curve in Fig.~\ref{fig_maincomp_theta0}. Specifically, in the range $\lambda_{\mathrm{K}}\lesssim 0.7$ (solid part), 
there are two types of stable solutions, that is, the KT solution (a stable limit cycle) and a solution corresponding to period-doubled orbits. In practice, however, we never observed these period-doubled orbits, indicating that their basin of attraction is much smaller than that of the KT solution.
At larger tumbling parameters ($\lambda_{\mathrm{K}}\gtrsim 0.7$), the PD1 bifurcation line becomes subcritical. This implies that the period-doubled solutions existing
for $\lambda_{\mathrm{K}}\gtrsim 0.7$ are unstable.
Instead, the solution evolves towards the next stable attractor, that is, the KT solution.

So far we have only considered the PD1 line, below which the T/W limit cycles are non-existent. 
 However, as we will discuss in more detail in the subsequent paragraph, the PD1 determines the transition between the T/W and the KT regime 
 only when the shear rate is swept from large to small values, that is, when
 the transition is approached from above. If, on the contrary, we increase the shear rate starting from a KT solution, this solution "dies"
 at a limit point of cycle (LPC). These points define the line labeled LPC in Fig.~\ref{fig_maincomp_theta0}. 
 To complete the discussion of KT states, we note that this type of solution can also be reached from the 
 flow-aligned (A) state, provided that the shear rate is not too large. This is seen in the right lower corner of the diagram Fig.~\ref{fig_maincomp_theta0}. The corresponding boundary curve
 is labeled as "LP2", where LP stands for "limit points" (saddle-node). 
 Crossing this line from the right, KT states are born via a "semi-global", that is, a SNIPER or SNIC bifurcation. A SNIC (Saddle-Node-on-an-Invariant Cycle) bifurcation results in the appearance of a limit cycle of an infinite period \cite{KuznetsovBook}. Indeed, upon approaching the LP2 curve from the left (i.e., increasing $\lambda_{\mathrm{K}}$) the period of the oscillatory motion
is found to increase rather steeply (corresponding to a slow-down of the tumbling frequency in a real system). Specifically, the period grows as $T\propto \abs{b-b_c}^{-1/2}$, where $b$ is the control parameter and $b_c$ is the critical parameter value at the SNIPER/SNIC bifurcation \cite{Broens1990,KaasPetersen1988}.
  
Another type of symmetry-breaking state is kayaking-wagging (KW), where the wagging motion of the director involves the space outside of the shear plane.
As seen from Fig.~\ref{fig_maincomp_theta0}, the KW state is stable at intermediate values of both, the tumbling parameter
and the shear rate. One way to reach the KW regime is to start from the W region (at suitable values of $\dot{\gamma}$, i.e. $2.4\lesssim \dot{\gamma}\lesssim 4.8$)
and to increase $\lambda_{\mathrm{K}}$.  Doing this, one crosses a line of branch points of cycles (BPC). Beyond this line, the (in-plane) wagging (W) ceases to exist, and a limit cycle
of type KW is born. We note that the appearance of BPCs implies that the KW oscillations are born in a pairwise manner. Thus, there is always a coexistence of two KW solutions in the parameter region labeled as KW in Fig.~\ref{fig_maincomp_theta0}. Upon increasing  $\lambda_{\mathrm{K}}$ even further, the KW disappears. As revealed from
Fig.~\ref{fig_maincomp_theta0}, this happens either at a second, subcritical period doubling curve (labeled PD2) or via a Neimarck-Sacker (NS) bifurcation. The latter leads
into a regime of chaotic solutions which are not further considered here (for a discussion of such states, see \cite{Rien02}). 
\subsubsection{Bistability}
\begin{figure}
\includegraphics[width=\columnwidth]{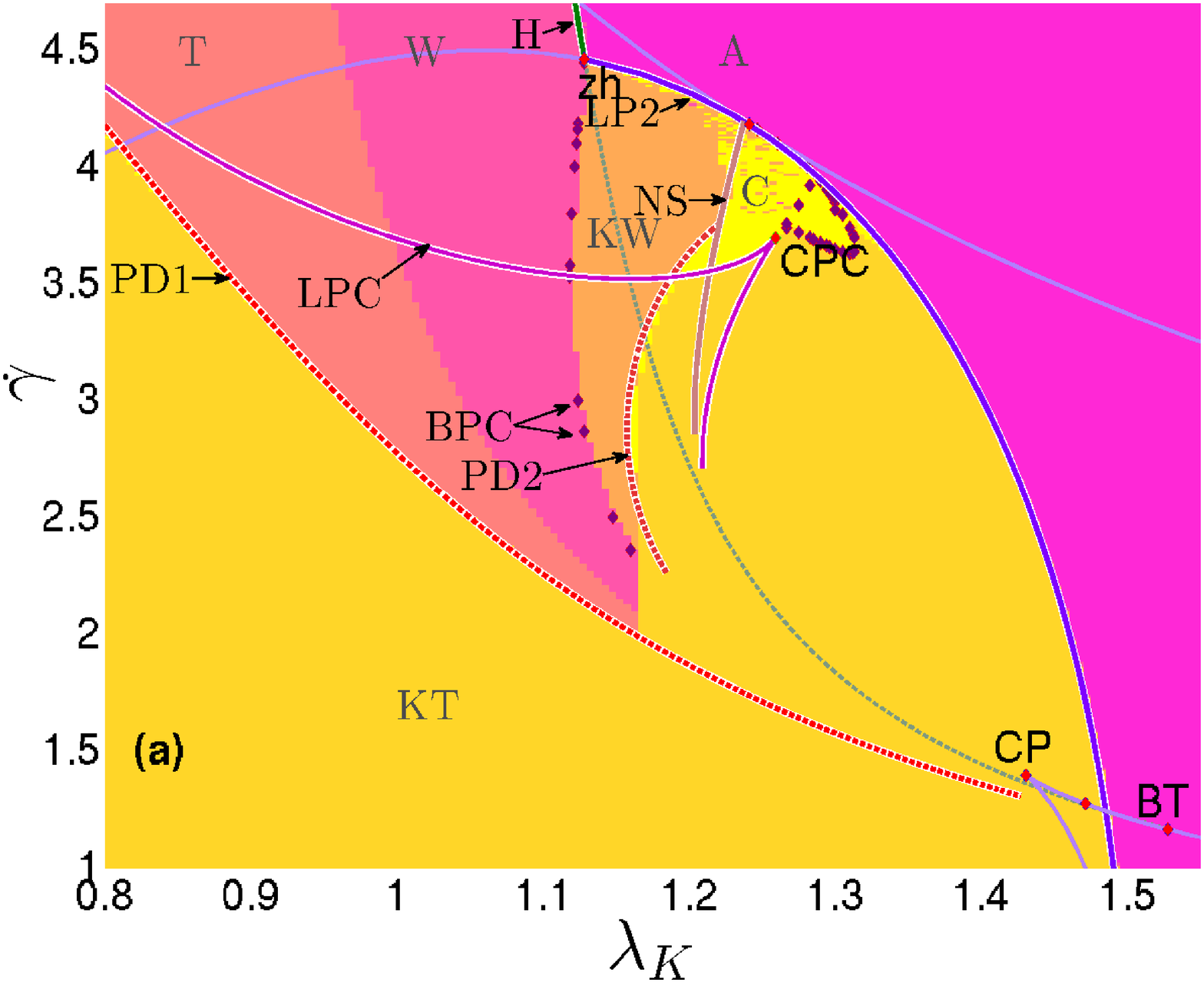}
\includegraphics[width=\columnwidth]{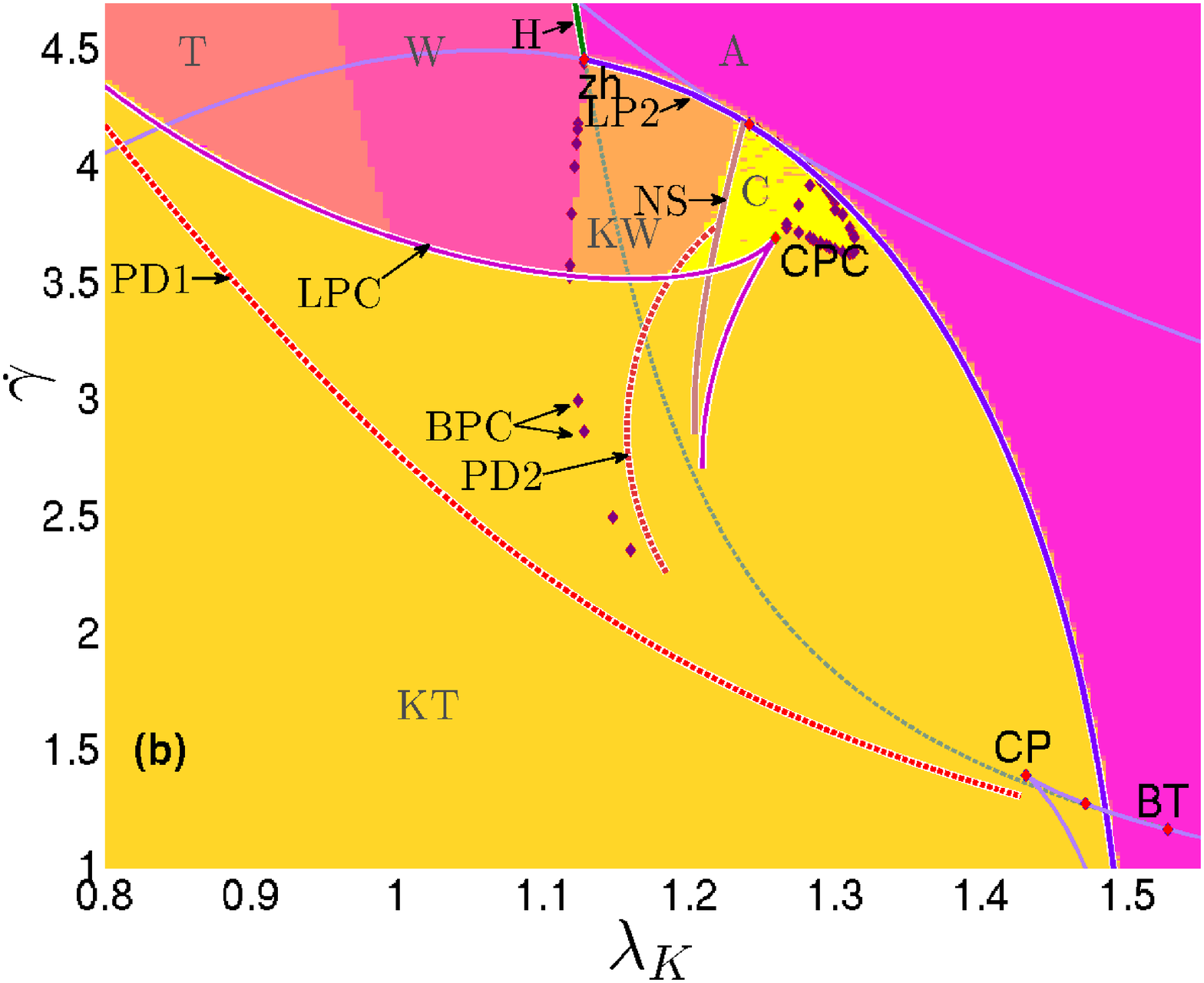}
\caption{\label{fig_bistability}
(Color online) Enlarged views of the parameter section of Fig.~\ref{fig_maincomp_theta0} in which bistability occurs.
The data in part (a) were obtained by sweeping the shear rate from higher to lower values, whereas in part (b), 
the shear rate was swept from lower to higher values. In each step, the final values of the integration is used as an initial condition for the
next integration step. To avoid staying on unstable solutions, the initial condition was perturbed by a small noise.}
\end{figure}

As already mentioned in the preceding paragraph, there are parameter regions where the dynamical behavior of the sheared system
is characterized by bistability (we do not further consider here the trivial coexistence of two KW states within the KW region)
To find bistable regions, we have investigated whether and how the results of the (direct) numerical integration
change when the parameters are varied in a different way. The resulting bistable regions are visualized in the two parts of Fig.~\ref{fig_bistability}.
The first region is bounded by the LPC line (from above), the dashed (subcritical) part of the PD1 line (from below), and the BPC line (from the right).
Depending on whether one enters that area from high or low shear rates, either in-plane T/W oscillations or out-of-plane KT oscillations are found.
The second region is again bounded by the LPC line from above, whereas the left and right boundaries
are the line of BPCs and the curve labeled PD2, respectively. Given the existence of these bistable regions it would be very interesting to see the consequences
of  this "coexistence" of dynamical states, if we allowed the system to be inhomogeneous. 
Indeed, first investigations in this direction \cite{Buran} already 
revealed the appearance of spatial domains characterized by different orientational states. These domains interact and change with time; an example being the
growth of one domain at the expense of another one. A detailed study of an inhomogeneous system is under way, but
is outside the scope of the present paper.
We note in this context that there are already some studies investigating inhomogeneous systems. In particular, 
Das \emph{et.~al} \cite{Das2005} studies, based on the same model we are using, inhomogeneous systems for similar shear rates and coupling parameters.
However, the main focus  was on detecting spatio-temporal chaos.
\subsection{Beyond the nematic phase} 
\label{subsec:HighT}
So far we have discussed the shear-induced nonlinear dynamics at the temperature $\theta=0$ corresponding to the {\it nematic} phase of the equilibrium system.
Given the variety of dynamic states observed at this low temperature, it is tempting to investigate the "fate" of these dynamical states when
$\theta$ is increased towards values 
where the equilibrium state becomes isotropic.
Clearly, under isotropic equilibrium conditions, any orientational ordering is {\it induced} by the shear flow. Indeed, previous theoretical investigations have already shown that
shear (applied to an originally isotropic system) can generate a flow-aligned state, paranematic state; moreover, it can shift the transition into the nematic state \cite{Olmsted92}.

However, usually this shear-induced, paranematic ordering is assumed to be {\it stationary}. This is also seen experimentally, e.g., for fd-viruses at concentrations below the
paranematic-nematic transition \cite{Lettinga05}. We note, however, that the particles involved in these experiments correspond to particular values of the
tumbling parameter,  $\lambda_{\mathrm{K}}$. For example, as described below Eq.~(\ref{tumb_shape}), fd-viruses are characterized by tumbling parameters of about
$\lambda_{\mathrm{K}}\approx 0.7-0.8$. There remains the question whether oscillatory dynamic states could exist 
beyond the nematic phase, if the particle's properties and thus, the tumbling parameter $\lambda_{\mathrm{K}}$ is changed.

\begin{figure}
 \includegraphics[width=\columnwidth]{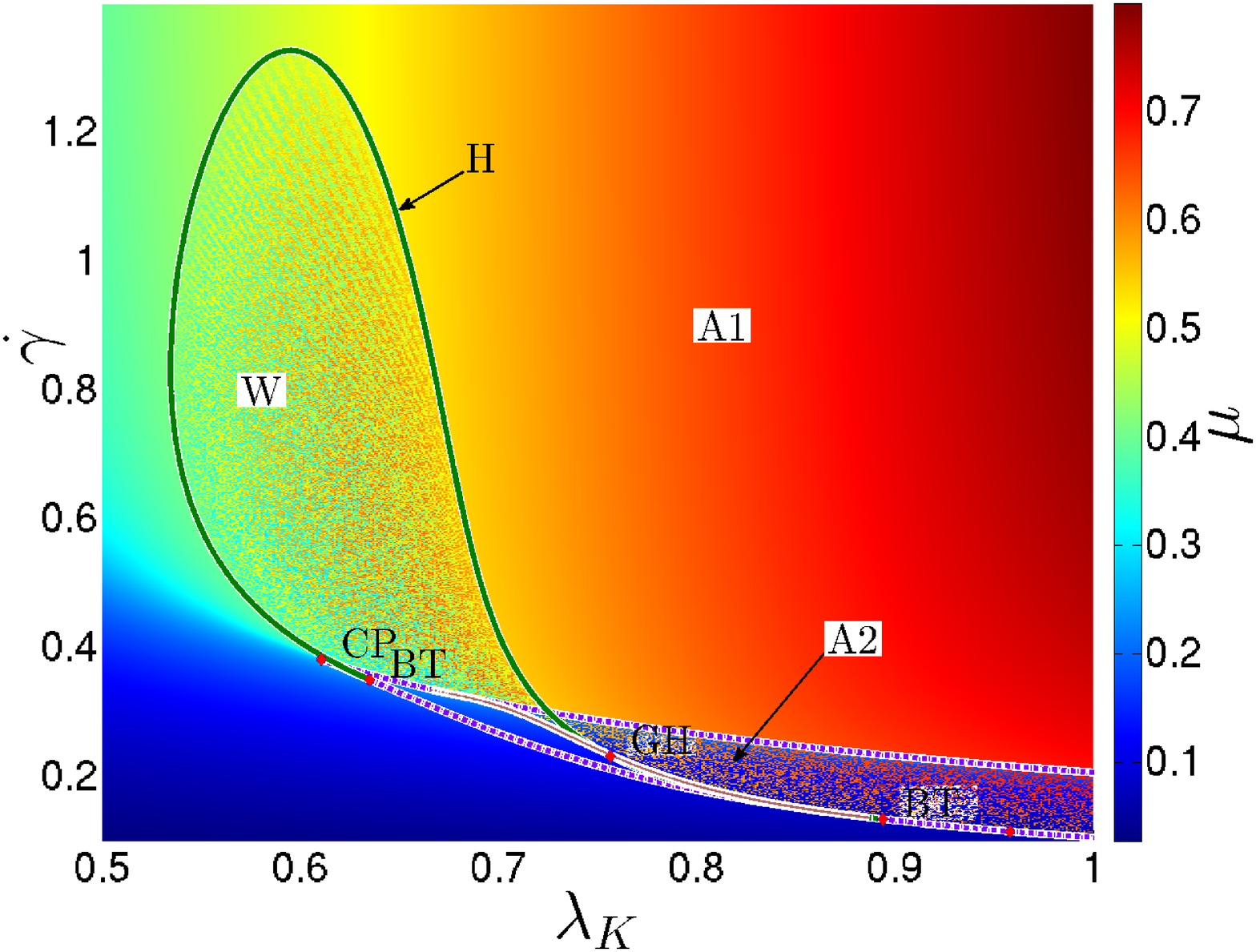}
\caption{\label{fig_overviewtheta120}
(Color online) State diagram at the reduced temperature
$\theta = 1.20$, where the equilibrium state ($\dot\gamma=0$) is isotropic.
The curves are obtained from a codimension-2-bifurcation analysis. The colors here reflect the value
of the eigenvalue $\mu=\mu_3$, which may be considered as
the degree of the uniaxial ordering (see color bar on the right side of the figure). One observes
an "island" of wagging (W) states. 
Region A1 corresponds to stable, shear-induced nematic order (flow-alignment). In region A2, there is a coexistence of a paranematic and a nematic state.
A closer view of this region (revealing even more states) is given in Fig.~\ref{fig_closetheta120}.
The thin red line (the dark gray vertical line at the bottom) indicates the path of the codimension-1 bifurcation, for fixed $\lambda_K=0.7$, that is discussed in Fig.~\ref{fig_codim1_lk07}.}
\end{figure}

To explore this issue within our mesoscopic approach we have extended the calculations at $\theta=0$ described in Sec.~\ref{subsec:NematicPhase} towards a range of larger values
of $\theta$, corresponding to higher real temperatures (smaller real concentrations) in thermotropic (lyotropic) systems.
An exemplary state diagram in the
$\lambda_{\mathrm{K}}$-$\dot\gamma$ plane is shown in Fig.~\ref{fig_overviewtheta120}. 
The reduced temperature is fixed to $\theta=1.20$, which is well above  the value $\theta=9/8=1.125$ (at $T_u$ or $c_u$, respectively) corresponding to the upper limit of the (metastable) nematic phase.
The curves and marked points have been obtained by MATCONT \cite{MATCONT}. In addition to these bifurcation lines, we have indicated in Fig.~\ref{fig_overviewtheta120}
the degree of nematic order induced by the shear flow. 
Specifically, the colors are chosen according to the value of the eigenvalue $\mu_3$ that characterizes the ordering in the direction of the director. 
In other words, the eigenvalue $\mu_3$ characterizes the degree of uniaxial ordering.
Note, however, that at nearly all parameter combinations shown in Fig.~\ref{fig_overviewtheta120} the orientational order is actually biaxial.

From Fig.~\ref{fig_overviewtheta120} we see that, at
small shear rates and tumbling parameters, the system is only weakly ordered, as expected at the high temperature (low concentration) considered. 
Increase of either $\lambda_{\mathrm{K}}$ or $\dot\gamma$ then leads to an increase of the order and hence larger $\mu_3$, reflecting shear-induced alignment. 

\begin{figure}
 \includegraphics[width=\columnwidth]{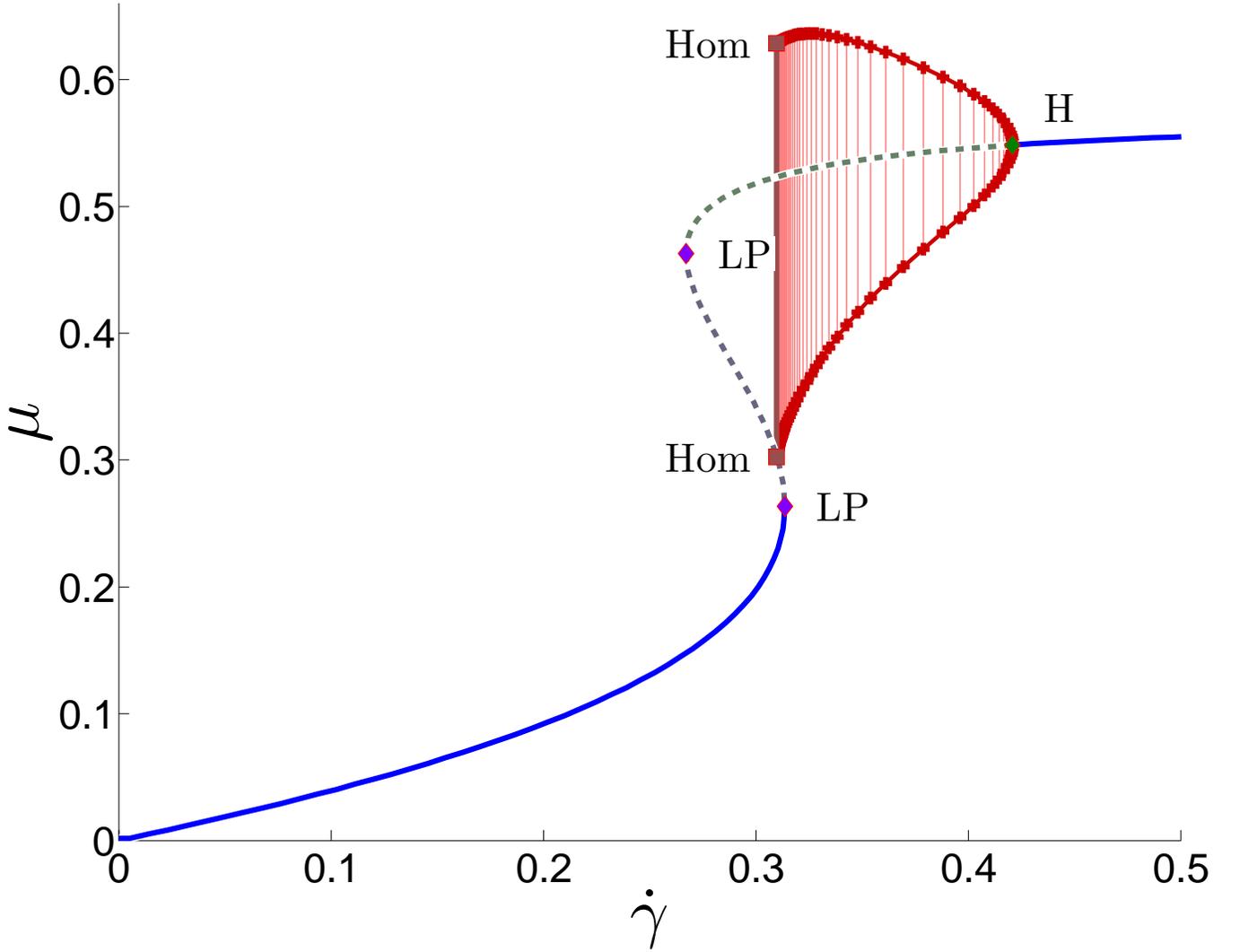}
\caption{\label{fig_codim1_lk07}
(Color online) Codimension-1 bifurcation diagram with the shear rate $\dot{\gamma}$ as control parameter and the largest eigenvalue $\mu=\mu_3$ of the alignment tensor as the order parameter.
The analysis is performed along the thin red line shown in Fig.~\ref{fig_overviewtheta120} and Fig.~\ref{fig_closetheta120} ($\theta=1.20, \lambda_K=0.7, \sigma=0$). The solid lines depict stable fixpoints. Dashed lines indicate unstable fixpoints.
Between the homoclinic bifurcation (Hom, colored brown [leftmost vertical line in the center]) and the supercritical Hopf bifurcation (H), there are stable limit cycles corresponding to the wagging state (W).
These limit cycles are illustrated by vertical lines between the maximal and minimal values of the limit cycles for the projection on the largest eigenvalue $\mu=\mu_3$.}
\end{figure}

However, the most important information from Fig.~\ref{fig_overviewtheta120} is that there is a range of shear rates and tumbling parameters where the shear induces {\it oscillatory} motion.
This parameter range is represented by the "island" in the left part of the figure, corresponding to tumbling parameters in the range
$\lambda_{\mathrm{K}}\lesssim 0.75$.
Within this island, the director performs a wagging (W) motion within the shear plane, implying that the degree of alignment oscillates in time. This is indicated
by the grainy color inside the W region. The grainy colors stem from different values of the largest eigenvalue $\mu_3$ during the oscillation. For every pair of parameters we determine $\mu_3$ only in one point of time. The upper boundary
of the oscillatory island is a Hopf bifurcation curve, whereas the lower boundary is a homoclinic orbit.\\
To illustrate the behavior of the largest eigenvalue $\mu_3$ of the alignment tensor as the order parameter versus the shear rate in this parameter region we provide a codimension-1 bifurcation analysis in Fig.~\ref{fig_codim1_lk07} for fixed $\lambda_K=0.7$ (see thin red line [vertical line at the bottom] in Fig.~\ref{fig_overviewtheta120}).
At the limit points (LP) and at the Hopf point (H) the stability of the fixpoint changes. The dashed lines indicate an unstable fixpoint.
The red lines depicts the minima and maxima of the limit cycle (in the projection to the largest eigenvalue $\mu_3$) that is evolving from the Hopf bifurcation point (H).
The period $T$ of the limit cycles tends to infinity, when the parameter is close to the homoclinic bifurcation (Hom) (colored brown [curve that connects BT and GH]).
The period grows logarithmically as the bifurcation is approached, i.e. $T\propto log(\abs{b-b_c})$, where $b$ is the control parameter and $b_c$ is the critical parameter value at the homoclinic bifurcation \cite{Broens1990,Gaspard1990,KaasPetersen1988}. Also note the hysteresis taking place in the small parameter region between the lower saddle-node bifurcation (LP) and the homoclinic bifurcation (Hom).
\\
To our knowledge, this is the first time that oscillatory states outside of the nematic region of the (equilibrium) phase diagram are found. 
One reason might be that most earlier theoretical studies based on the present, mesoscopic approach \cite{Das2005,Hess2004} focus on
larger values of the tumbling parameter. Indeed, in the range $\lambda_{\mathrm{K}}\gtrsim 0.75$
the present calculations reproduce the (well-established) fact of a shear-induced, first-order, paranematic-nematic transition \cite{Hess2004,Olmsted92}.
Consider, as an example, the case $\lambda_{\mathrm{K}}=0.8$. Increasing the shear rate from small values,  the system remains essentially disordered (as indicated by the blue color [dark color in the bottom]) up to
$\dot\gamma\approx 0.2$. At this point, one crosses the LP bifurcation line and enters a bistable region characterized by two stable fixed points, that correspond
to a paranematic ordering coexisting with a (stationary) 
shear-aligned state. In a plot of the order parameter versus shear rate (not shown here), this bistable region
(labeled A2 in Fig.~\ref{fig_overviewtheta120})
would correspond to the hysteresis region. Finally, upon increasing $\dot\gamma$ even further, the pseudo nematic ordering becomes unstable and one enters
the region A1. Here, the order parameter is relatively large, reflecting pronounced shear-alignment. 

\begin{figure}
 \includegraphics[width=\columnwidth]{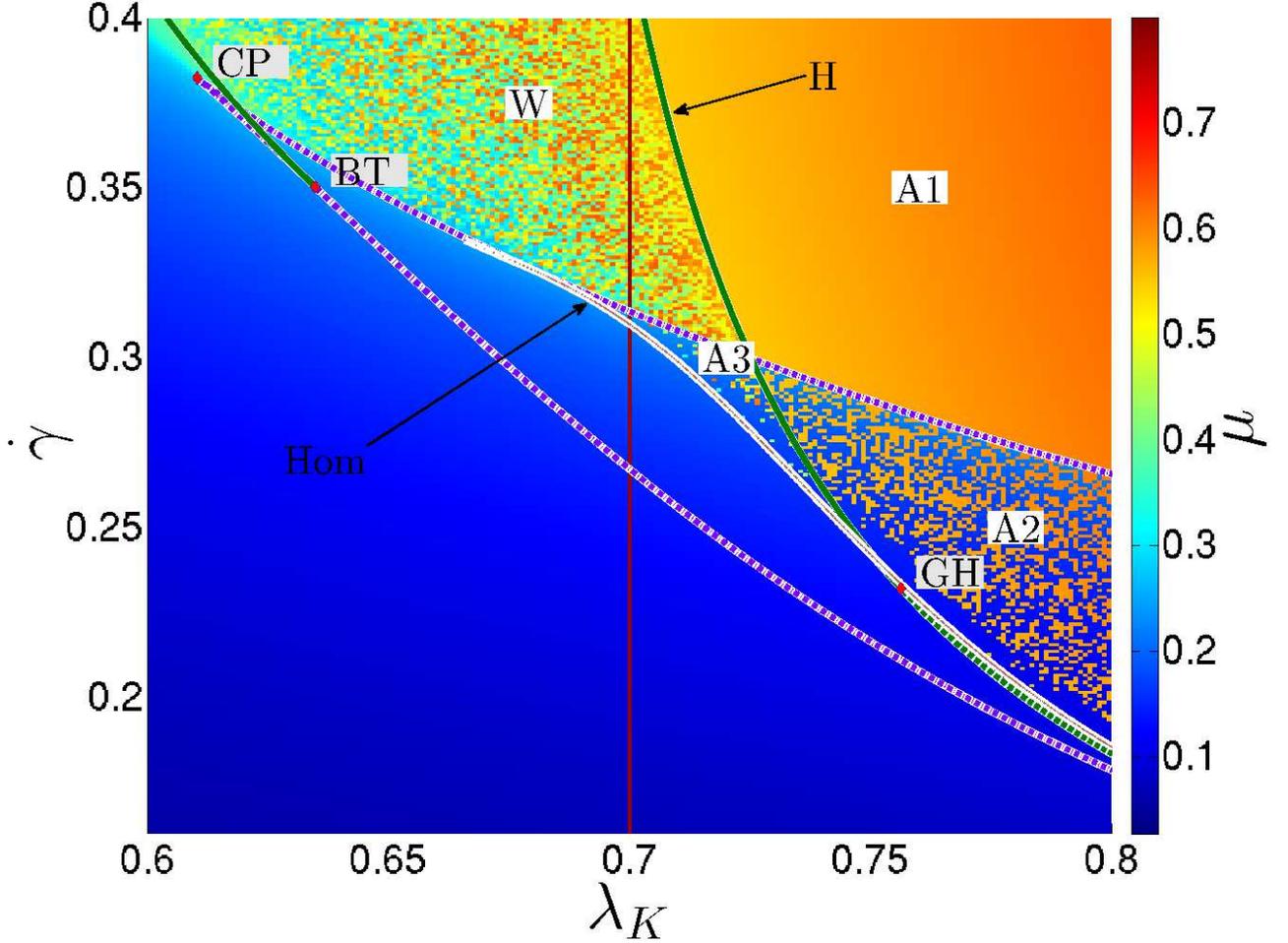}
\caption{\label{fig_closetheta120}
(Color online) Enlarged view of the parameter section of Fig.~\ref{fig_overviewtheta120} at the border of the wagging island and the A2/A1 regions
($\theta = 1.20$). Within the region labeled A3, stationary alignment (with low value of the order parameter) coexists 
with an oscillatory (W) solution.}
\end{figure}

The bifurcation scenario occurring at the parameters where the wagging island, on the one hand, and the A2/A1 region, on the other hand, approach one another,
is rather complicated. This is illustrated in Fig.~\ref{fig_closetheta120}, which shows an enlarged view of the relevant section of 
Fig.~\ref{fig_overviewtheta120}.

 In particular, one can see that there is a small, additional region labeled A3. In this region one finds both, a stable fixed
point representing weak flow-alignment, and a stable limit cycle representing dynamical wagging. In other words, A3 is again a bistable region. The upper boundary of this region
is either a limit point (LP) saddle-node bifurcation curve (smaller values of $\lambda_{\mathrm{K}}$) or a Hopf bifurcation curve (larger values of $\lambda_{\mathrm{K}}$).
Finally, the lower boundary is a homoclinic curve. Below the homoclinic curve the wagging solutions does not exist any more.
\subsection{Non-equilibrium phase diagrams} 
\label{overview}
In the preceding paragraph we have studied the system in the parameter plane spanned by the shear rate, $\dot\gamma$, and the coupling parameter $\lambda_{\mathrm{K}}$, the latter being related to the shape of the particles [see Eq.~(\ref{tumb_shape})]. Thus, it is clear that $\lambda_{\mathrm{K}}$  is not an adjustable quantity in a specific, experimental system. Here, the relevant parameters are rather the temperature or concentration, respectively, and the shear rate or its conjugate, the shear stress, while the tumbling parameter $\lambda_{\mathrm{K}}$
is constant (up to polydispersity effects).

\begin{figure}
 \includegraphics[width=\columnwidth]{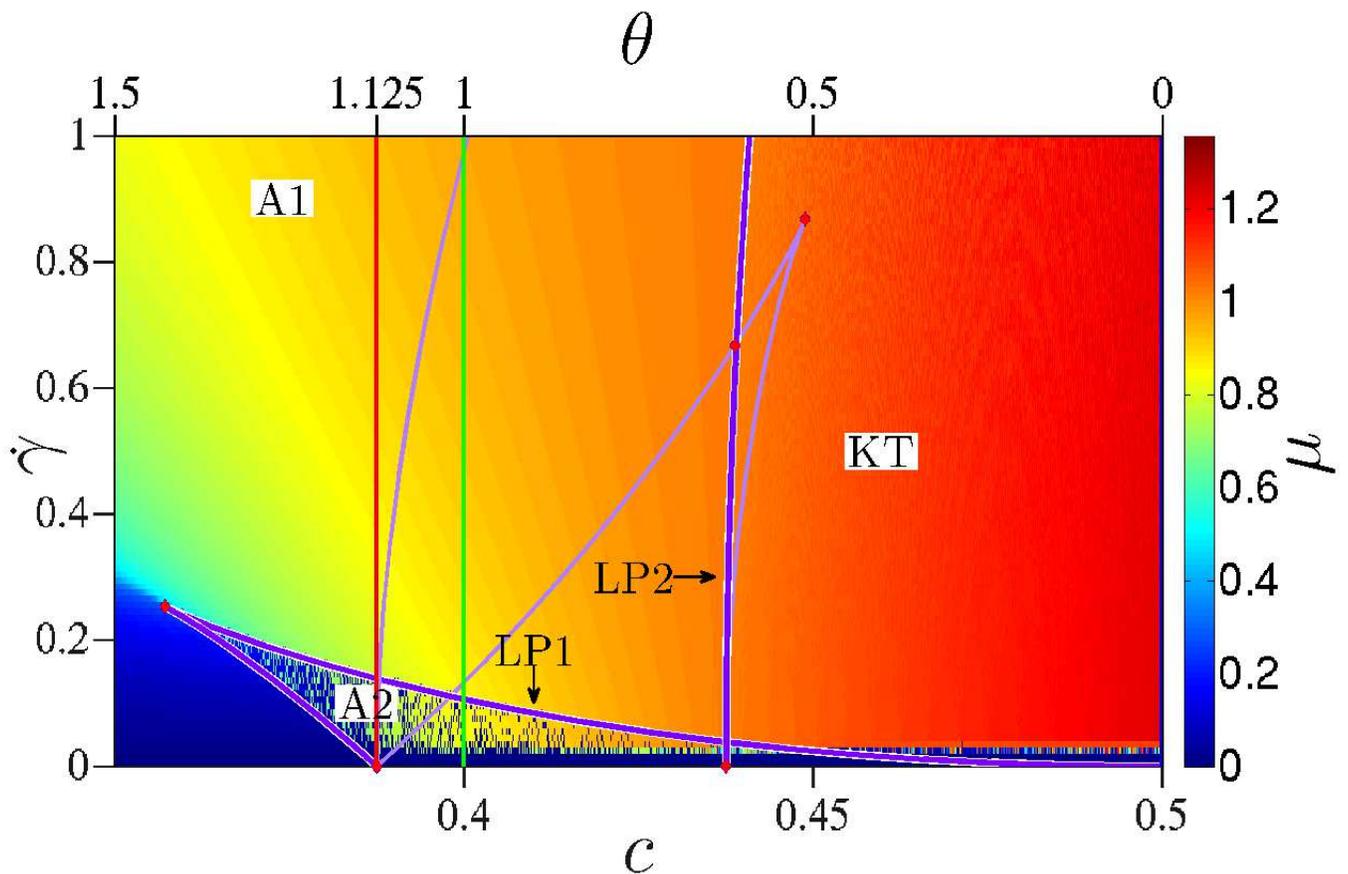}
\caption{\label{fig_thetalk125new}
(Color online) Non-equilibrium phase diagram in the plane spanned by the concentration (bottom horizontal axis) or temperature (top horizontal axis) and the shear rate (vertical axis) at 
fixed tumbling parameter $\lambda_{\mathrm{K}}=1.25$. 
The colors represent the eigenvalue $\mu=\mu_3$ of the alignment tensor $\bm{a}$. We also show the bifurcation lines obtained by a codimension-2 analysis.
In the triangular area labeled A2, we find bistability between a paranematic and a nematic state with flow-alignment.}
\end{figure}

\begin{figure}
 \includegraphics[width=\columnwidth]{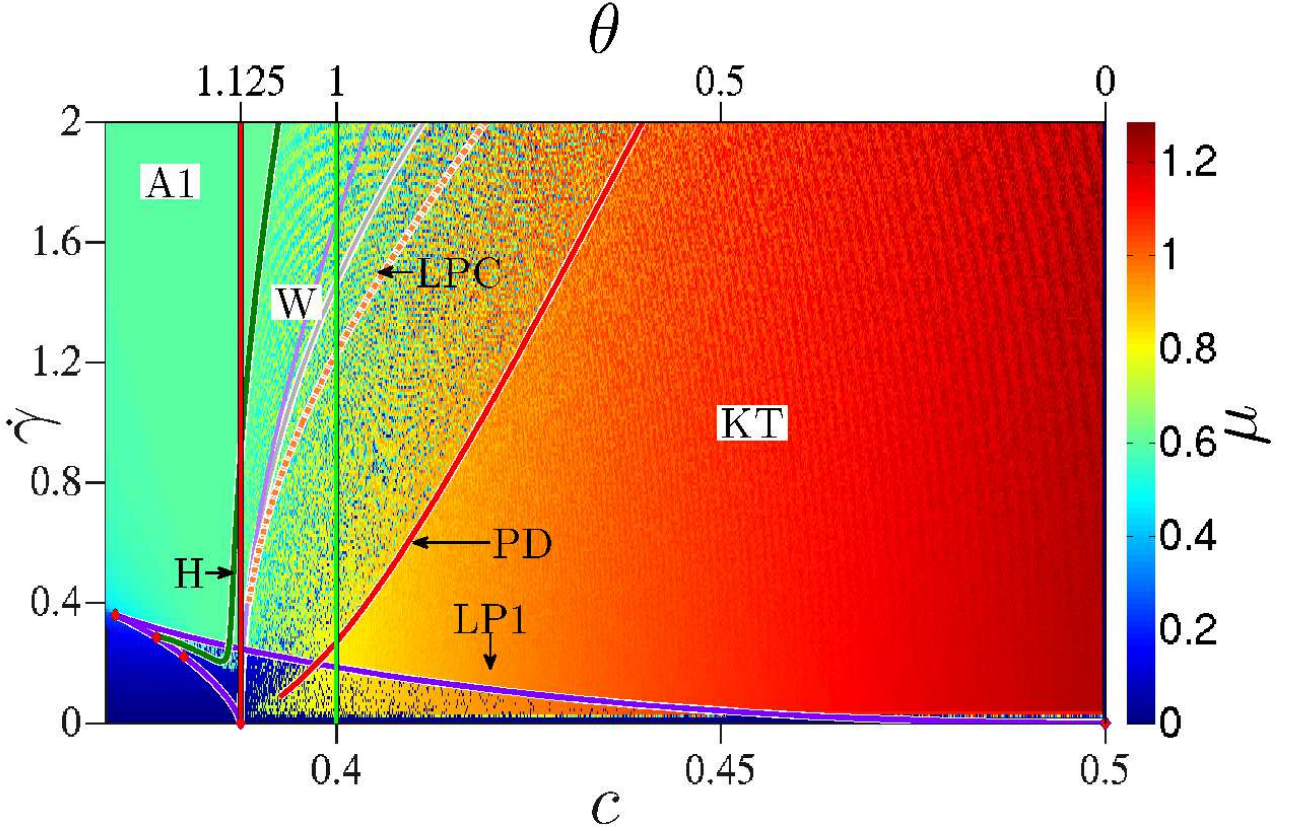}
\caption{\label{fig_thetalk074}
(Color online) Same as Fig.~\ref{fig_thetalk125new}, but for $\lambda_{\mathrm{K}}=0.74$.
In the area right of the Hopf bifurcation curve (H) we observe oscillatory states. Specifically, these are
Wagging/Tumbling (W/T) in the area between the Hopf (H) and the period doubling (PD) bifurcation curve, and Kayaking/Tumbling (KT) 
in the area right of the PD curve. In between the LPC and the PD curve, we find bistability between these states.}
\end{figure}

\begin{figure}
\includegraphics[width=\columnwidth]{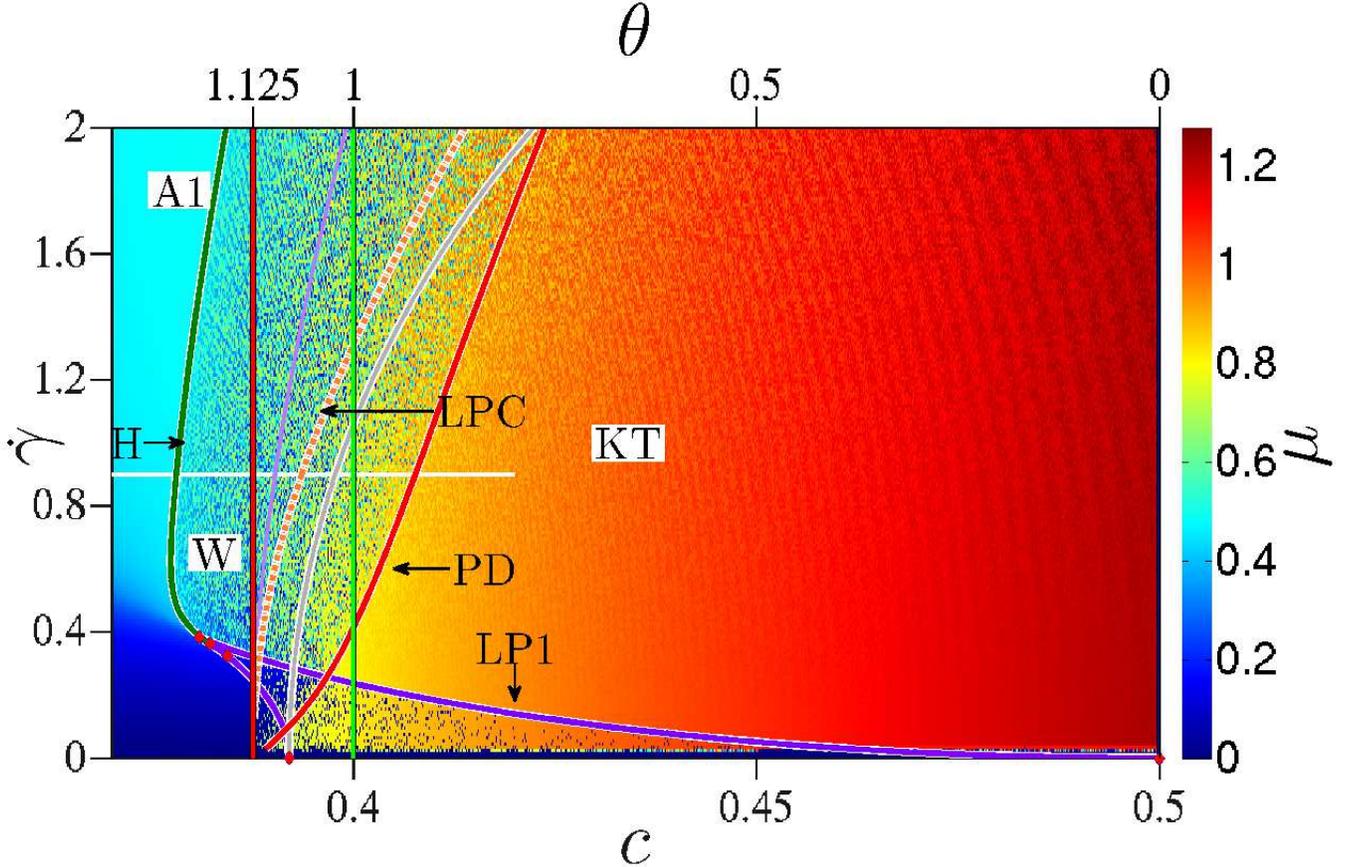}
\caption{\label{fig_thetalk06new}
(Color online) Non-equilibrium phase diagram for $\lambda_{\mathrm{K}}=0.6$. Compared to Fig.~\ref{fig_thetalk074} ($\lambda_{\mathrm{K}}=0.74$), the wagging region (W) is more extended towards the concentration (temperature) region, where the equilibrium state is isotropic (i.e. left from the red [leftmost] vertical line).
The thin white horizontal line indicates the path for the codimension-1 diagram presented in Fig.~\ref{fig_thetalk06_Codim1}.}
\end{figure}

Motivated by this experimental situation we present in Figs.~\ref{fig_thetalk125new}-\ref{fig_thetalk06new}
state diagrams in the $\theta$-$\dot\gamma$ or $c$-$\dot\gamma$ plane. These diagrams may be considered as non-equilibrium phase diagrams
(as a generalization of the equilibrium phase diagrams corresponding to the case $\dot\gamma=0$). We have obtained these diagrams
for three values of the tumbling parameter $\lambda_{\mathrm{K}}$, that is, at $\lambda_{\mathrm{K}}=1.25$, $0.74$, and $0.6$. 
These values seem particularly interesting judging from our previous analysis in 
Secs.~\ref{subsec:NematicPhase}-\ref{subsec:HighT}. 

In the subsequent figures~\ref{fig_thetalk125new}-\ref{fig_thetalk06new}, the lower $x$-axis shows the concentration, $c$,
while the upper $x$-axis shows additionally the reduced temperature. The $y$-axis denotes the shear rate. 
By choosing $c$ as a main variable, we take into account the fact that many recent experimental studies \cite{Lettinga05,Lettinga2004} focus on lyotropic systems. 
To map the quantities $c$ and $\theta$ onto each other we have employed the second member of Eq.~(\ref{theta}) that can be
solved with respect to $c$. Clearly, application of the resulting equation requires an input for the clearing point concentration, $c_{\mathrm{K}}$, and the pseudo-critical concentration, $c^{*}$.
Here we have arbitrarily chosen $c_{\mathrm{K}} = 0.4$ and $c^{*} = 0.5$.

In Fig.~\ref{fig_thetalk125new},
we have indicated $c_{\mathrm{K}}$ (corresponding to $\theta=1$) and $c^{*}$ ($\theta=0$) by colored vertical lines. In addition, we have put a vertical line at the concentration 
$c_u=0.3875$ ($\theta=9/8=1.125$) below (above) which the nematic state is absolutely unstable. Finally, 
the color of the different regions in Fig.~\ref{fig_thetalk125new} indicate the degree of nematic order [see Eq.~(\ref{order_parameter})], neglecting biaxiality. 
Thus,  the lower $x$-axis corresponds to the
equilibrium phase diagram of a lyotropic system.

We start by considering the case $\lambda_{\mathrm{K}} = 1.25$ illustrated in Fig.~\ref{fig_thetalk125new}. 
At small concentrations below the stability limit of the nematic phase (i.e., left from the red [leftmost] vertical line) and small, yet non-zero, values of $\dot\gamma$ we observe 
a paranematic state with very weak nematic order. Increasing the shear rate at a fixed $c$ in this range of concentrations, the system enters first a region (labeled A2)
of bistability between the paranematic and a shear-aligned state, and finally the true shear-aligned state (A1). These phenomena are characteristic of the well-known, shear-induced paranematic-nematic transition
already mentioned in Sec.~\ref{subsec:HighT}. Indeed, following earlier studies we can interpret the two violet lines labeled by LP1 as "spinodals" of the isotropic-nematic transition under shear and their merging point ($c_{\mathrm{c}}$, $\gamma_{\mathrm{c}}$)
as "critical point" of this transition. The analogy to an equilibrium critical point (of, e.g. the vapor-liquid transition of a fluid) is that at shear rates above the critical point, the transition becomes continuous, that is, the order parameter increases smoothly when $c$ is increases at fixed  $\dot\gamma>\dot\gamma_{\mathrm{c}}$.
Consistent with earlier studies \cite{Lenstra2001,Olmsted92}, we find that the spinodal lines are curved towards the left. This is plausible, since shear tends to enhance the degree of nematic ordering and thus supports the transition into a (shear-)aligned state.
We note that similar effects in the non-equilibrium phase diagram have been observed in studies of the impact of biaxial stretching flow \cite{Rey1995}.
\\
We also find from Fig.~\ref{fig_thetalk125new} that at $\lambda_{\mathrm{K}} = 1.25$, oscillatory states only appear at high concentrations (low $\theta$) within the nematic regime. Consistent with our analysis of the case $\theta=0$/$c=c^{*}$ in Sec.~\ref{subsec:NematicPhase}, the states are of kayaking-tumbling (KT) type in the range of shear rates considered (compare to Fig.~\ref{fig_maincomp_theta0}). 

In Sec.~\ref{subsec:HighT} we have found that at smaller values of $\lambda_{\mathrm{K}}$, oscillatory states can become stable even at reduced temperatures or concentrations
where the equilibrium system is isotropic. Motivated by this observation, we present in Figs.~\ref{fig_thetalk074} and \ref{fig_thetalk06new} $c$-$\dot\gamma$ diagrams at
$\lambda_{\mathrm{K}} = 0.74$ and $\lambda_{\mathrm{K}} = 0.6$. 


\begin{figure}
\includegraphics[width=\columnwidth]{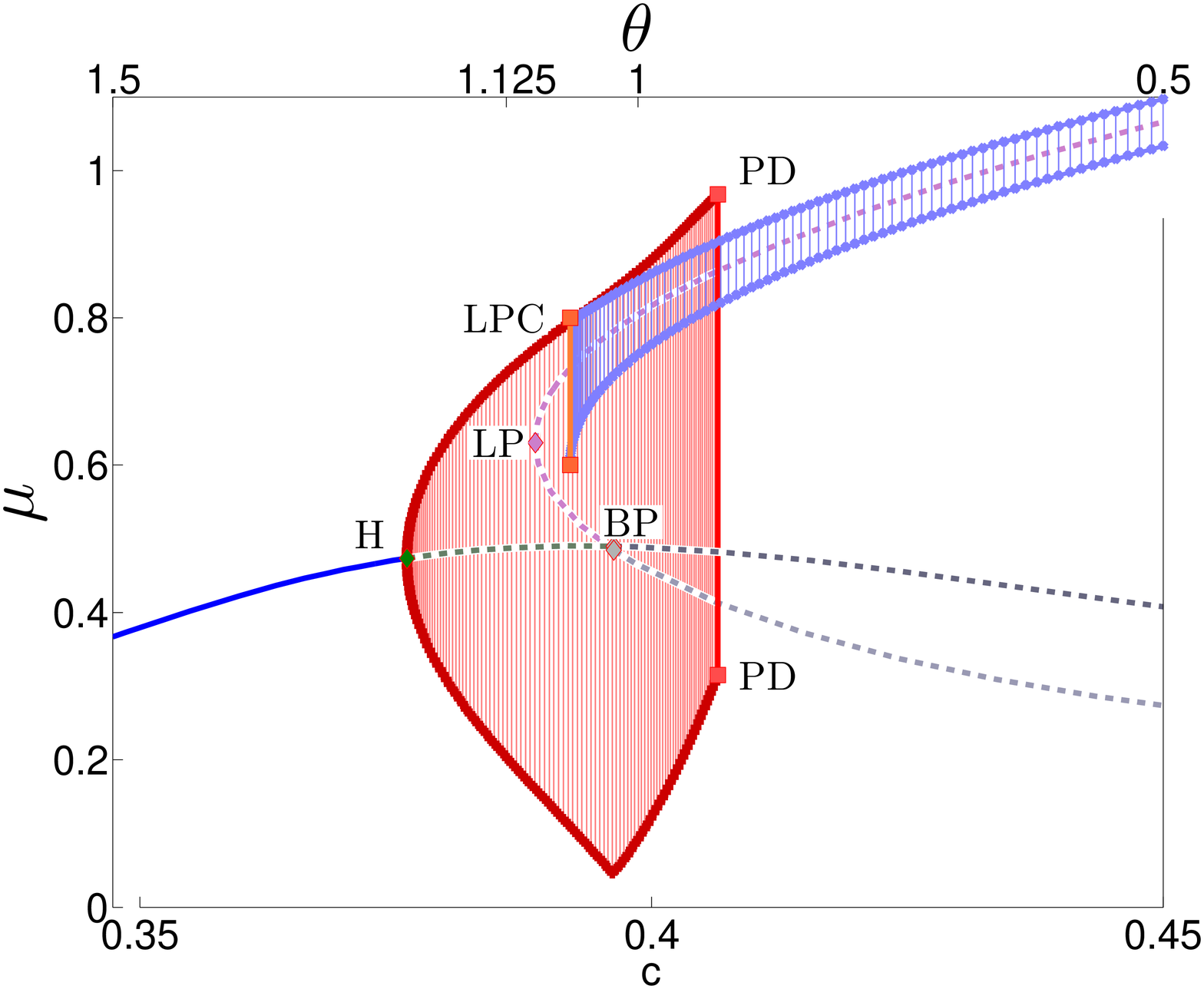}
\caption{\label{fig_thetalk06_Codim1}
(Color online) Codimension-1 diagram using the bifurcation parameter $c$ ($\theta$).
The path of the bifurcation analysis is along the white horizontal line in Fig.~\ref{fig_thetalk06new} ($\dot{\gamma}=0.9, \lambda_{\mathrm{K}}=0.6, \sigma=0$).
Solid (dashed) lines indicate stable (unstable) fixpoints. The limit cycles are illustrated by vertical lines between the maximal
and minimal values for the projection on the largest eigenvalue $\mu=\mu_3$ as the order parameter. Note the bistable region of wagging states (red [dark gray] limit cycles) and kayaking tumbling
states (blue [light gray] limit cycles) between the period doubling bifurcation (PD) and the limit point of cycle (LPC) bifurcation.}
\end{figure}
Similar to the case of $\lambda_{\mathrm{K}} = 1.25$, one may identify the spinodal of the paranematic-nematic transition, with the "critical" shear rate  
$\dot\gamma_{\mathrm{c}}$
being somewhat larger than before. This mean that higher shear rates are necessary to see a complete disappearance of the first-order transition.
However, the most important new feature appearing at lower values of $\lambda_{\mathrm{K}}$ is that the Hopf (H) curve, which marks the onset of oscillatory states, moves 
(at least partially) into the parameter region left of the red, vertical line, that is, into the regime where the equilibrium nematic state is absolutely unstable. These oscillatory states 
are of type wagging (W) or tumbling (T) [as argued before, there is no substantial difference between these motions from the perspective of the bifurcation analysis].
Increasing the concentration further, systems at $\lambda_{\mathrm{K}} =0.74$ or $\lambda_{\mathrm{K}} =0.6$ 
first enter a region where, depending on initial conditions, the dynamics is either of type W/T or of type KT. This region is located
in between the LPC curve (limit point of cycle) and the PD (period doubling) curve. Finally, at concentrations beyond the PD curve, only T states are stable.

Coming back to the regime of low concentrations we see from Fig.~\ref{fig_thetalk074} that at 
$\lambda_{\mathrm{K}}=0.74$, the oscillatory region left of the (nematic) stability limit is rather small; in fact, the major part of the corresponding Hopf line lies inside the spinodal. Interpreted in terms of a real system, this would mean that oscillations do occur, but only at concentrations larger than the critical
one (defined by the top of the spinodal), i.e., inside the true nematic phase of the sheared system.
This is indeed consistent with recent experimental results for fd-viruses under shear flow.
In these experiments, the full paranematic-nematic binodal at shear rates $\dot{\gamma} > 0$ was mapped out.
The ``coexisting'' nematic phase is characterized by tumbling motion, whereas the paranematic state is flow-aligned. The tumbling-to-aligning transition line then ends at the maximum of the binodal \cite{Ripoll2008,Ripoll2008a,DhontChapter3GompperSchickVol2}.

The behavior just discussed changes at even smaller values of the tumbling parameter such as the case $\lambda_{\mathrm{K}} =0.6$. Indeed it is seen from Fig.~\ref{fig_thetalk06new} that,
upon increasing $\dot\gamma$ from zero in the relevant parameter region (i.e., left of the red [leftmost], vertical line), the Hopf curve starts only at the critical point $(\dot\gamma_{\mathrm{c}},c_{\mathrm{c}})$ and then initially bends to the {\it left}. In other words, W or T states can exist already at the less ordered side of the
paranematic-nematic transition, consistent with the $\dot\gamma$-$\lambda_{\mathrm{K}}$ diagram discussed in
Sec.~\ref{subsec:HighT} [see Fig.~\ref{fig_overviewtheta120}]. The corresponding behavior of the order parameter $\mu_3$, i.e., the largest eigenvalue of the alignment tensor, as function of $c$ ($\theta$) is shown in Fig.~\ref{fig_thetalk06_Codim1}.
Only at larger values of the shear rate we finally observe a bending of the Hopf curve towards higher concentrations. This latter behavior may be interpreted such that strong shear typically favors shear-aligned states (A1) rather than oscillatory motion (compare with Fig.1).

\section{Conclusions}
In the present study we have employed a numerical path continuation analysis combined with mesoscopic equations of motion
to investigate the complex dynamical behavior of a homogeneous system of rod-like particles under planar 
(Couette) shear flow. Compared to a conventional integration of the mesoscopic equations \cite{GrandnerEPJ2007,GrandnerPRE2007,Heidenreich2008Thesis,Rienacker2000}, 
a main advantage of the continuation method is that it provides not only full information about the (long-time) dynamics at different system parameters, but also
about the {\em nature} of the bifurcation lines separating different regions of the non-equilibrium phase diagram. This information is not only of theoretical, but also of practical interest.
For example, close to the bifurcation line separating different states, the continuation method
can predict the behavior of the amplitude and frequency of an order parameters when one crosses the bifurcation line by varying a suitable system parameter (e.g., shear rate, concentration, \dots). Furthermore, the method proved to be very helpful in determining multistable regions, that might otherwise be missed due to their small extension in parameter space.
On top of these insights, however, we have found a new aspect of the non-equilibrium behavior which was formerly unknown. Namely, the sheared system can exhibit an oscillatory
state (wagging) at temperatures/concentrations {\em outside} the nematic region of the equilibrium phase diagram. 
To our knowledge this is the first time not only in theory, but also from the experimental perspective that oscillatory states outside the nematic region have been observed. One possible
reason might be that, according to our results, the oscillatory states only occur for rather small values of the coupling parameter $\lambda_{K}$, corresponding to rod-like particles
with relatively small aspect ratio. Real systems such as suspensions of fd-viruses typically involve more elongated particles, for which our theory predicts oscillatory states only 
{\em within} the nematic region, consistent with experiments \cite{Ripoll2008a}. Indeed, our non-equilibrium phase diagram (obtained by performing calculations for a large range of concentrations) has strong similarities with that presented in Ref.~\cite{DhontChapter3GompperSchickVol2}.

Based on the present results one can now proceed towards the investigation of inhomogeneous systems, including the question how different spatial regions interact
within the multistable regions of the parameter space. A particularly interesting aspect in this context concerns the front speeds \cite{Loeber2012,Ebeling1991} and the basins of attraction
of the involved states. Another motivation of including inhomogeneities is to explore the appearance of "banded" states, such as vorticity banding (which has already been observed in experiments \cite{Kang2006}) and gradient banding \cite{DhontChapter3GompperSchickVol2}. Work in these directions is in progress.

\appendix
\section{}
\label{appA}
{In this Appendix we present some considerations regarding the structure and the symmetries of Eqs.~(\ref{eq:HomSyst1}-\ref{eq:HomSyst2}). To start with, we rewrite the right-hand sides of these equations. The Landau-de Gennes free energy in Eq.~(\ref{LG}) can be written in projections of the tensorial basis \cite{Rien02a} as
\begin{align}
  {\Phi} = {\frac{1}{2}}({\theta}+a^2)a^2-a_0\left[a_0^2-3(a_1^2+a_2^2)+{\frac{3}{2}}(a_3^2+a_4^2)\right]\notag\\
-{\frac{3}{2}}{\sqrt{3}} \left[a_1(a_3^2-a_4^2)+2a_2a_3a_4 \right].
\label{eq:LG:basis}
\end{align}
In Eqs.~(\ref{eq:LG:basis}), $a^2 =\sum_{i=0}^4 a^2_i$.\\
We now consider several cases.
\\
i) For vanishing shear rate $\dot{\gamma}=0$ the dynamic equations reduce to
\begin{align}
\label{eq:gradient}
  {\frac{d}{dt}}a_i= -{\Phi_i}= -{\frac{{\partial}}{{\partial}a_i}}{\Phi}.
\end{align}
The $\Phi_i$ are identical to those appearing in Eqs.~(\ref{eq:HomSyst2}). From Eq.~(\ref{eq:gradient}) we see that the  alignment follows
a gradient dynamics. Therefore, no oscillatory solutions are possible for $\dot{\gamma}=0$, i.e., oscillatory solutions are all flow-induced \cite{Strogatz_1994}.
To see this, we determine the Jacobian $\bm{J}$ of the right hand side of Eq.~(\ref{eq:gradient}) as
\begin{align}
\label{jacobian}
 \bm{J}_{ij} = \frac{\partial}{\partial a_j} (- \Phi_i) & = -  \frac{\partial}{\partial a_j} ( \frac{\partial}{\partial a_i} \Phi) = -
  \frac{\partial}{\partial a_i} ( \frac{\partial}{\partial a_j} \Phi) = \bm{J}_{ji}.
\end{align}\\
The Jacobian $\bm{J}_{ij}$ is symmetric due to the symmetry of the second-order derivatives. A symmetric matrix has only real eigenvalues. Thus, when performing a linear stability analysis based on $\bm{J}$, we cannot find any oscillations.\\
ii) For the case $\dot{\gamma}\neq0$ but $\sigma=0$, it is instructive to introduce a function H defined as
\begin{align}
  H ={\frac{{\dot{\gamma}}}{2}}(a_1^2+a_2^2+{\frac{1}{2}}(a_3^2+a_4^2)-{\sqrt{3}}{\lambda}_Ka_1).
\end{align}
Note that H is independent of $a_0$ and proportional to the shear rate $\dot{\gamma}$. With this function, Eqs.~(\ref{eq:HomSyst1}-\ref{eq:HomSyst2}) can be rewritten as
\begin{align}
\label{eq:Heq}
  {\frac{{\mathrm{d}}}{{\mathrm{d}}t}}a_0 & =-{\frac{{\partial}}{{\partial}a_0}}\Phi \notag\\
  {\frac{{\mathrm{d}}}{{\mathrm{d}}t}}a_1 & =-{\frac{{\partial}}{{\partial}a_1}}\Phi+{\frac{{\partial}}{{\partial}a_2}}H\notag\\
  {\frac{{\mathrm{d}}}{{\mathrm{d}}t}}a_2 & =-{\frac{{\partial}}{{\partial}a_2}}\Phi-{\frac{{\partial}}{{\partial}a_1}}H\notag\\
  {\frac{{\mathrm{d}}}{{\mathrm{d}}t}}a_3 & =-{\frac{{\partial}}{{\partial}a_3}}\Phi+{\frac{{\partial}}{{\partial}a_4}}H\notag\\
  {\frac{{\mathrm{d}}}{{\mathrm{d}}t}}a_4 & =-{\frac{{\partial}}{{\partial}a_4}}\Phi-{\frac{{\partial}}{{\partial}a_3}}H
\end{align}
The structure of the Eqs.~(\ref{eq:Heq}) reflects the presence of two oscillatory subunits ($a_1$,$a_2$) and ($a_3$,$a_4$). The partial derivative of $H$ with respect to $a_2$ appears in the dynamical equation for $a_1$ and vice versa the partial derivative of $H$ with respect to $a_1$ in the dynamical equation for $a_2$. 
Similarly, the partial derivative of $H$ with respect to $a_4$ appears in the dynamical equation for $a_3$ and vice versa the partial derivative of $H$ with respect to $a_3$ in the dynamical equation for $a_4$.
In particular, oscillations in $a_1,a_2$ with $a_3=a_4=0$ indicate tumbling or wagging within the shear plane. If there are additionally oscillations in $a_3$ and $a_4$ we have out of shear plane kayaking states.\\
iii) In order to make a similar reformulation of the dynamics for the more general case $\dot{\gamma}\neq0$ and $\sigma\neq0$, two additional dissipative couplings between $a_0$ and $a_2$ as well as between $a_3$ and $a_4$ must be added. In this spirit, we amend the potential $\Phi$ in Eqs.~(\ref{eq:Heq}) by an additional term $V$ (i.e., $\Phi\longrightarrow\Phi+V$). 
For the amendment we have
\begin{align}
  V = {\sigma}{\dot{\gamma}}({\frac{1}{2}}a_3a_4-{\frac{1}{3}}{\sqrt{3}}a_0a_2).
\end{align}
We note that the additional term $V$ for $\sigma\neq0$ only contributes to the relaxational part of the dynamic equation.
\\
So in general the Eqs.~(\ref{eq:HomSyst1}-\ref{eq:HomSyst2}) can be expressed as derivatives of $\Phi,V$ and $H$.
\section{}
\label{appB}
In this Appendix we give a short description on how we determined our bifurcation diagrams for homogeneous systems based on the freely available software package MATCONT \cite{MATCONT}.
As shown in Sec. \ref{subsec:eqmotionhomogeneous}, the present dynamical system can be formulated as ${d\bm{x}}/{dt}=\bm{f}(\bm{{x}},\bm{\beta})$, where the vector $\bm{x}$ contains the components $a_0$,$a_1$,\dots,$a_4$ and $\bm{\beta}$ contains the bifurcation parameters.
Here we are dealing with a codimension-2 diagram, where one has two bifurcation parameters (all other parameters are fixed). We investigate mainly two cases. In the first case the bifurcation parameters are $\lambda_{\mathrm{K}}$ and $\dot{\gamma}$, whereas the dimensionless temperature $\theta$ [see Eq.~(\ref{theta})] is kept fixed.

If we now choose a set of parameters $\bm{\beta}=\bm{\beta_0}$ and start to integrate the system with initial conditions $\bm{{x}}=\bm{{x_{0}}}$, there are three possibilities for the outcome of this procedure (disregarding transients).
First, the solution goes to a stable fixed point (equilibrium point/EP); second, it goes to a stable periodic oscillatory solution (limit cycle/LC) or third, it goes to a possibly irregular and chaotic attractor.
From a practical point of view, it is most convenient to choose $\bm{x_0}$ and $\bm{\beta_0}$ such that one ends up in an EP or LC solution. Starting from that solution one changes one parameter, i.e., one performs a codimension-1 bifurcation analysis.

The continuer algorithm then tries to stay on the EP or LC solution, while also monitoring the Jacobian of the system and its eigenvalues.
Typically one can encounter saddle node or limit point bifurcation (LP), a Hopf bifurcation H, a limit point of cycles (LPC), a torus or Neimarck-Sacker bifurcation (NS), and a flip or period doubling bifurcation (PD) (see \cite{MATCONT2}).
These codimension-1 bifurcations points are then used as starting points for a codimension-2 bifurcation analysis. In this way one can calculate branches of LP, H, LPC, NS or PD depending on two bifurcation parameters.
Following these branches one can also encounter special codimension-2 points like a Cusp point (CP), Bogdanov-Takens point (BT), Zero Hopf (ZH) or Generalized Hopf (GH). 
We note that this list is by no means complete.

A characteristic feature of codimension-2 points is that they correspond to an intersection of different branches of bifurcations. For example, a BT corresponds to the intersection of a Hopf branch, a LP branch and a homoclinic branch. Based on that knowledge one can now attempt to continue along the missing branches.
Reiteration of these steps finally yields the full codimension-2 bifurcation diagram.

We are noting that, within our bifurcation analysis, both KT and KW solutions are indeed rather hard to find. The reason is that these symmetry-breaking states
are not born in local bifurcations (such as the Hopf bifurcation separating A and W states) but rather in {\it global} bifurcations. To identify such a bifurcation
it is not sufficient to just monitor the Jacobian and its eigenvalues. However, once one has obtained a limit cycle solution of type KW or KT, the numerical continuation
via MATCONT allows to follow these solutions in parameter space, until the boundary is reached.\\

\begin{acknowledgments}
We gratefully acknowledge financial support from the Deutsche
Forschungsgemeinschaft through the Research Training Group GRK~1558 (projects B.3 and C.4). We also thank Sebastian Heidenreich for fruitful discussions.
\end{acknowledgments}

\bibliography{Strehober.bib}

\end{document}